**Post-growth annealing of GaMnAs under As capping - an alternative way to increase Tc**

M. Adell (1 and 3), V. Stanciu (2), J. Kanski (1), L. Ilver (1), J. Sadowski (3 and 4), J. Z. Domagala (4), P. Svedlindh (2), F. Terki (5), C. Hernandez (5), S. Charar (5)
((1) Department of Experimental Physics, Chalmers University of Technology, SE-41296 Goteborg, Sweden, (2) Department of Engineering Sciences, Uppsala University, Box 534, SE-751 21 Uppsala Sweden, (3) MAX-lab, Lund University, Box 118, SE-221 00, Sweden, (4) Institute of Physics, Polish Academy of Sciences, al . Lotnikow 32/46, PL-02-668 Warszawa, Poland, (5) Groupe d'Etude des Semiconducteurs CC074, Universite Montpellier II, France)

We demonstrate that *in situ* post-growth annealing of GaMnAs layers under As capping is adequate for achieving high Curie temperatures ($T_C$) in a similar way as *ex situ* annealing in air or in $N_2$ atmosphere practiced earlier. Thus, the first efforts give an increase of $T_C$ from 68 K to 145 K after 2 hours annealing at 180 °C. These data, in combination with lattice parameter determinations and photoemission results, show that the As capping acts as an efficient sink for diffusing Mn interstitials.

Following the discovery of ferromagnetism in Mn doped III-V semiconductors[1], $(III_{1-x}Mn_x)V$, with paramagnetic-to-ferromagnetic transition temperatures, $(T_c)$ in excess of 100 K, there has been an immense interest in magnetic properties of diluted magnetic semiconductors (DMS). In recent years, considerable progress has been done both in understanding the material properties, and improving the quality of low-temperature molecular beam epitaxy (LT MBE) grown GaMnAs films[2-5]. This has led to an increase of $T_c$ in GaMnAs from the previously "established" limit[6] of 110 K to 160 – 170 K, as is now reported by several research groups[4,7,8]. The increase of $T_c$ by 50 – 60 K was possible due to the identification and understanding of the role of structural defects, which compensate the Mn acceptors in GaMnAs. Since GaMnAs can only be grown by LT MBE, it hosts all the defects typical to LT GaAs[9-11]. At the MBE growth conditions suitable for GaMnAs deposition the prevailing structural defect is the As antisite ($As_{Ga}$). It is well known[10-12] that the maximum concentration of $As_{Ga}$ is close to 0.5%. Similar or even higher concentration of $As_{Ga}$ is expected in GaMnAs[13,14]. In addition there is one defect specific to GaMnAs, namely Mn in interstitial positions ($Mn_I$). The importance of this defect was first analyzed theoretically[15], and then $Mn_I$ was identified experimentally[16]. The negative impact of $Mn_I$ on the ferromagnetic properties of GaMnAs is due to two effects: $Mn_I$ acts as a double donor, and thus tends to compensate the p-doping induced by Mn in substitutional sites. Furthermore, the magnetic moment of Mn in interstitial sites tends to be antiferromagnetically coordinated to the magnetic moment of Mn in substitutional sites, thus reducing the average magnetic moment per Mn atom[17]. It is now generally accepted that the main effect of post-growth annealing, leading to a remarkable increase of Tc, is the removal of interstitial Mn. The detailed mechanism responsible for this was discussed recently by Edmonds et al[18]. The key feature is out-diffusion of Mn interstitials towards the GaMnAs surface and their passivation. Passivation of $Mn_I$ is usually achieved by the oxidation or nitridation - all successful post-growth annealing experiments reported so far were performed ex-situ, after exposing the GaMnAs samples to air and annealing either in $N_2$ atmosphere[2,19] or in air[20]. The best ferromagnetic- and transport properties in GaMnAs are obtained for very thin layers (100 – 1000 Å), which appears to be a natural consequence of an increased impact of the surface on the overall layer properties. For very thin films (50 Å and below) the thickness of the oxidized surface region may be comparable to that of the original GaMnAs layer. In this case the layer thickness, essential in the analysis of Hall effect data, is not well known.

In this letter we analyze the magnetic, structural and electronic properties of (Ga,Mn)As that is annealed under an As capping layer. Since the As capping protects the surface from oxidation, the annealing can be carried out in air as well as in vacuum. We show that there is no essential difference in the annealing efficiency between the two environments. The behavior of the As cap during annealing in vacuum was monitored by reflection high energy electron diffraction (RHEED). The annealing temperature in vacuum was measured with an infrared pyrometer, and in air by means of a thermocouple in contact with the Mo-block holding the In-glued (Ga,Mn)As samples.

We have grown a set of $Ga_{0.94}Mn_{0.06}As$ layers with thicknesses in the range of 100 Å to 1000 Å, keeping the MBE growth procedure the same for all samples. Before sample growth, the Ga and Mn fluxes were carefully calibrated using RHEED intensity oscillations from a separate test sample. The GaMnAs samples were grown on semi-insulating GaAs(100) epi-ready substrates. Samples intended for photoemission were grown on n-type (Te doped) substrates. Each growth started with deposition of a 1000 Å thick high temperature (HT) GaAs buffer, and all samples were grown at the maximum substrate temperature (230 °C) allowing accommodation of 6% Mn. To avoid interfacial n-p profiles, which are believed to degrade the transport properties of thin GaMnAs films[21], no LT GaAs buffer was deposited. The (Ga,Mn)As growth was monitored by RHEED intensity oscillations. Usually it was possible to observe oscillations of the specular RHEED beam throughout the whole growth. After (Ga,Mn)As deposition the As effusion cell shutter and As source valve were closed (an $As_2$ valved cracker source was used) and the substrate heater was shut down. When the substrate temperature dropped below 100 °C the As valve and the As shutter were re-opened and a thick amorphous As capping layer was deposited. We estimate the thickness of this As layer to be in the range of 2000 – 5000 Å. At the beginning of As deposition the typical streaky RHEED pattern, reflecting a smooth and well-ordered (Ga,Mn)As surface, changed to a diffuse intensity distribution of the amorphous As layer.

After As deposition the (Ga,Mn)As sample was either annealed in the MBE growth chamber (with an As background pressure of $10^{-9}$ mbar), or taken out of the vacuum chamber and cleaved to smaller pieces for alternative annealing treatments. One of them was put back to the MBE growth chamber, where it was subjected to LT annealing, other pieces were annealed in air, and one was kept as a reference.

Figure 1 shows results from magnetization vs. temperature measurements for three As-capped $Ga_{0.94}Mn_{0.06}As$ samples with thicknesses 400 Å, 700 Å and 1000 Å, denoted s1, s2, and s3, respectively. Pieces of s1 were annealed at 180 °C in air for 1, 2, 3, and 6 hrs, while pieces of s2 and s3 were annealed for 1, 2, 3, and 30 hrs in air at the same temperature (180 °C). It can be seen that an optimum annealing time at this temperature is reached quite quickly and is close to 2hrs regardless of the sample thickness. The highest $T_c$ value is obtained for the 400 Å thick sample. The increase of $T_c$ from 68 K in the as-grown film to 145 K (see Fig. 1b) after "optimum" annealing, i.e. 77 K, is the highest reported so far to the best of our knowledge. The variation of saturation magnetization with respect to the annealing time follows the same pattern (see inset Fig. 1c). It is also seen that after passing the 2hrs peak in $T_c$, extended annealing at 180 °C results in slightly reduced $T_c$. As expected, there is no significant difference between the results of annealing in air and in vacuum - the amorphous As layer protecting the (Ga,Mn)As surface is not desorbed upon annealing at 180 °C for times shorter than 12hrs (as verified by RHEED). Desorption of the amorphous As capping layer could be achieved, however, by extended annealing at 230 °C, and this was used to remove the capping for subsequent photoemission investigations of the annealed surface. Thus, the (Ga,Mn)As/As interface was the same in samples annealed in vacuum and in air.

In Fig.2 we show the effects of annealing on the resistivity of a 300 Å thick $Ga_{0.94}Mn_{0.06}As$ sample (s4). The 1 h long post-growth annealing leads to significant reduction of the resistivity (by about 50%), as well, as shifts of R(T) maximum towards higher temperature. This latter effect is associated with the corresponding increase of $T_c$ value.

In Fig. 3, we show the influence of annealing on the lattice parameter of the 700 Å thick $Ga_{0.94}Mn_{0.06}As$ layer (s2). As reported before[13, 22], one effect of post-growth annealing is a reduction of the (Ga,Mn)As lattice parameter. This is interpreted as an effect of removing $Mn_I$ atoms from the (Ga,Mn)As volume. Both experimental and theoretical studies[13, 22, 23] suggest that Mn interstitials are responsible for the observed lattice parameter increase in GaMnAs with Mn content, while substitutional Mn in Ga sites has only a small effect in this respect. The GaMnAs lattice parameter values calculated from angular positions of (004) Bragg diffraction peaks measured for as-grown as well as annealed pieces piece of s2 are given in Table 1.

Thus, our results clearly indicate that Mn interstitials are removed from the (GaMn)As lattice by annealing under an As capping., in a similar way as by annealing in air or in nitrogen atmosphere, though on a shorter time scale than reported by Edmonds et. al.[2, 18]. The mobility of Mn atoms in the GaMNAs lattice is not likely to be affected by the surface conditions, so we ascribe the depletion of Mn interstitials to a very efficient trapping of diffusing Mn atoms by the As capping. Reaching the surface, these atoms can bind chemically with As to form MnAs. Since MnAs is a metallic compound, one would then expect the surface to exhibit metallic properties. Indeed, photoemission data clearly demonstrate that this is the case, as shown in Fig. 4: after annealing under As capping a well-defined Fermi step is observed when the capping is removed. In contrast, without As capping the surface remains semiconducting after a similar annealing process. The photoemission results (to be discussed in greater detail in a separate paper) also show that the magnitude of the Fermi step is proportional to the thickness of the GaMnAs layer (for thicknesses up to 1000 Å). This implies that the metallic layer accommodates Mn atoms from the whole (GaMn)As layer. The detailed structure of this metallic surface remains to be determined, but the available experimental information strongly suggests that it consists of epitaxial MnAs: apart from the already mentioned metallicity, we know that the surface is flat (the RHEED pattern remains streaky), it is well ordered (a clear 1x2 surface reconstruction is seen in both RHEED and LEED), and the work function is uniform and well defined (as reflected by a sharp low-energy cut-off of the photoemission spectra). Furthermore, preliminary observations of the integer order LEED spot intensities indicate clear differences between the (GaMn)As surface and the surface obtained after As decapping. The two surfaces are thus basically different, even though their in-plane geometries are the same.

The inferred growth of zinc-blende MnAs must of course be verified by other techniques, especially as growth of MnAs/GaAs digital alloy structures by vapor phase epitaxy shows that zinc-blende MnAs can only be grown in the form of submonolayer islands[24,25] and theory predicts MnAs to be unstable in this structure[26].

In conclusion, we have shown that an amorphous As cap provides an efficient sink for Mn interstitial atoms diffusing during low temperature post-growth annealing. The presence of an amorphous As cap protects the (GaMn)As layer from the ambient atmosphere, which means that the annealing can be carried out in air as well as in vacuum. An advantage of the present method is that it allows epitaxial overgrowth of the annealed (GaMn)As by e.g. GaAs. In this way the present preparation method opens the possibility to incorporate the magnetic layers in epitaxial multilayer structures. It should be mentioned that similar results are obtained by an amorphous Sb cap, but since the desorption temperature of Sb is significantly higher, As capping is preferred in situations where epitaxial overgrowth is desired. The method avoids surface oxidation, thus excluding the influence on the magnetic properties of the annealed GaMnAs samples by unidentified magnetic phases such as Mn oxides or nitrides. On the other hand, the MnAs layer formed during annealing may have a corresponding undesirable effect. However, considering the magnetic properties discussed here, we stress that a MnAs film in the monolayer range cannot contribute significantly to the measured magnetization of GaMnAs films thicker than 50 Å.

**Acknowledgements**

The present work is part of a project supported by the Swedish Research Council (VR) and Swedish Foundation of Strategic Research (SSF). One of the authors (J.S) acknowledges the financial support form the Polish State Committee for Scientific Research (KBN) through the grants No: 2-P03B-05423 and PBZ-KBN-044/P03/2001. The authors would like to thank Dr M. Sawicki from the Institute of Physics, Polish Academy of Sciences, Warszawa, Poland for valuable discussions.

**Figure captions**

**Fig.1.** (a) - dependence of paramagnetic-to-ferromagnetic phase transition temperature of three amorphous As capped $Ga_{0.94}Mn_{0.06}As$ films with thicknesses of 400Å, 700Å, 1000Å on the annealing time; (b) – temperature dependence of magnetization for sample s1, after 2hrs post-growth annealing; (c) – saturation magnetization vs. post-growth annealing time for sample s1

**Fig. 2.** Temperature dependence of resistivity of amorphous As capped 300Å thick $Ga_{0.94}Mn_{0.06}As$ layer. Significant reduction of resistivity, as well as $T_c$ increase manifested by shift of R(T) maximum towards higher temperatures can be seen.

**Fig.3.** X-ray diffraction results - 2θ/ω scans, (004) Bragg reflection for 700 Å thick $Ga_{0.94}Mn_{0.06}As$ layer (sample s2). Right side peak – Bragg reflection from GaAs(001) substrate; left side peaks – Bragg reflection from GaMnAs layer. The annealing induced GaMnAs lattice parameter reduction upon annealing can clearly be seen.

**Fig.4.** Normal emission valence band photoelectron spectra from as-grown and As-capping-annealed $(Ga_{0.94}Mn_{0.06})As$. The spectra were excited with 81 eV photons. The bottom curve shows the difference between the two upper spectra, and represents essentially the metallic surface layer.

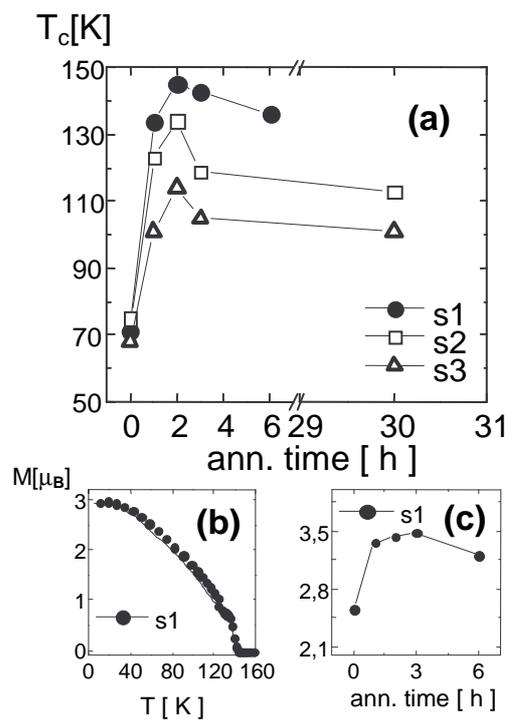

*M. Adell et. al. Fig.1*

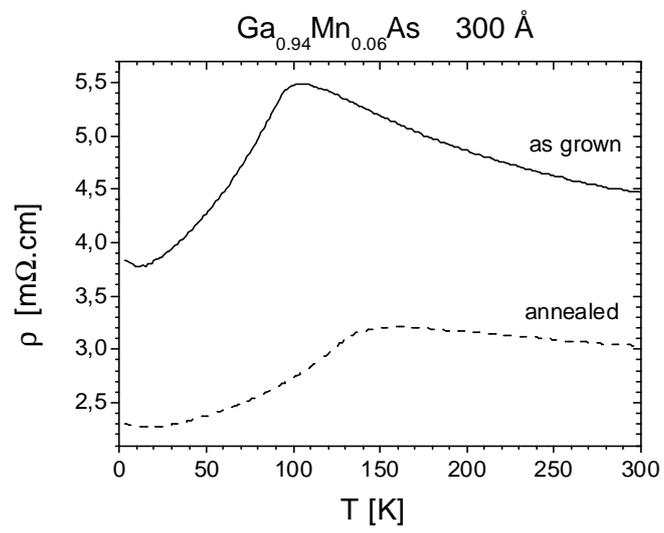

*M. Adell et. al. Fig.2*

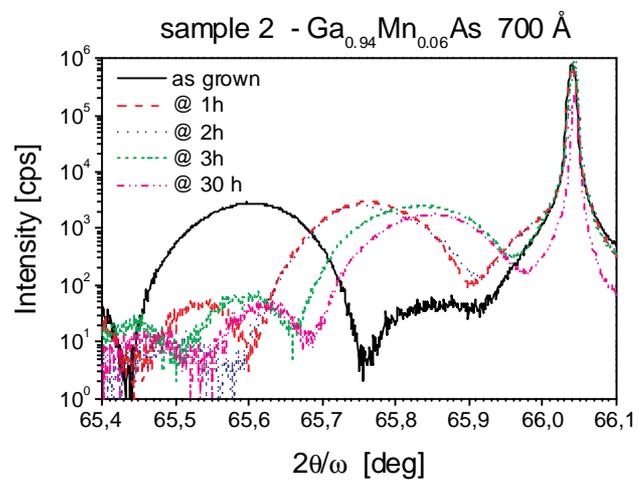

M. Adell et. al. Fig.3

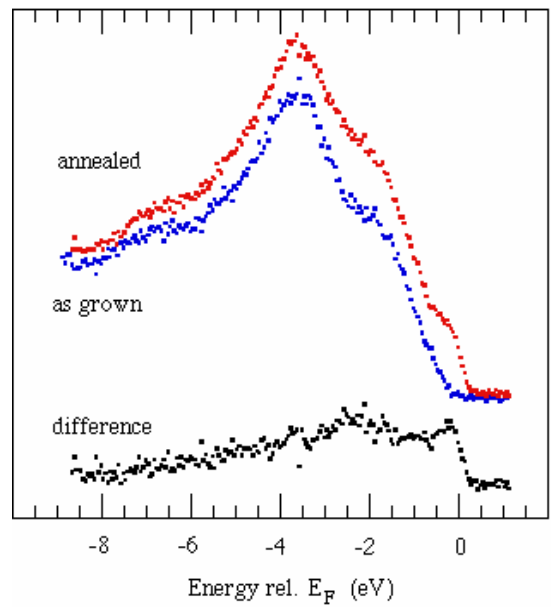

*M. Adell et. al. Fig.4*

**Table 1.** Lattice parameter of the 700 Å thick $Ga_{0.94}Mn_{0.06}As$ film (s2) before and after post-growth annealing. Parallel lattice constant is always the same and is not changed upon annealing. Significant reduction of the perpendicular lattice parameter after the post-growth annealing has been observed.

| Sample | $a_\parallel$ [Å] | $a_\perp$ strained [Å] | $a_\perp$ relaxed [Å] | $\delta a/a$ |
|---|---|---|---|---|
| as grown | 5.65348 | 5.68727 | 5.669947 | 3055 |
| annealed 1h | 5.65348 | 5.67533 | 5.664128 | 1978 |
| annealed 2h | 5.65348 | 5.67506 | 5.663997 | 1953 |
| annealed 3h | 5.65348 | 5.67075 | 5.661896 | 1564 |
| annealed 30h | 5.65348 | 5.66795 | 5.660532 | 1311 |